\documentclass[11pt]{article}
\date{}
\usepackage[dvips]{epsfig}
\usepackage{amsthm}
\usepackage{amsmath}
\usepackage{amssymb}
\newcommand{\ot}{{\,\otimes\,}}
\newcommand{{\Cd}}{{\mathbb{C}^3}}

\def\<{\langle}
\def\>{\rangle}

\begin{document}

\title{\bf A family of generalized Horodecki-like entangled states}


\author{Dariusz Chru\'{s}ci\'{n}ski and Adam Rutkowski\\
 Institute of Physics, Nicolaus Copernicus University \\
 Grudzi\c{a}dzka 5/7, 87--100 Toru\'{n}, Poland }

\maketitle

\begin{abstract}
We provide a multi-parameter family of 2-qudit PPT entangled states
which generalizes the celebrated Horodecki state in $3 \ot 3$. The
entanglement of this family is identified via semidefinite
programming based on``PPT symmetric extensions" by Doherty et al.
\end{abstract}

\section{Introduction}

The problem to determine whether a given quantum state is separable
or entangled, is one of the most fundamental problems in
Entanglement Theory \cite{HHHH}. Starting from the famous
Peres-Horodecki PPT (Positive Partial Transpose) criterion
\cite{Peres}, nowadays there are enormous number of different
separability criteria  (see e.g. \cite{O1,O2,O3} and
\cite{HHHH,Guhne} for the recent reviews). It turn out that among
known separability criteria, those based on ``symmetric extensions
and ``PPT symmetric extensions, developed by Doherty et al.
\cite{Doherty-1,Doherty-2} are considered to be the most effective.
It turns out that both NPT and PPT symmetrically extendable states
can be characterized by semidefinite programming, a well-known
optimization problem for which many free solvers are available (like
the MATLAB toolbox SeDuMi \cite{SEDUMI}). For the recent approach to
symmetric extensions see also \cite{Plenio}. In the present Letter
we use these criteria to identify entanglement of the new class of
PPT states in $\mathbb{C}^d \ot \mathbb{C}^d$. This family provide
the multi-parameter generalization of the seminal Horodecki state in
$\mathbb{C}^3 \ot \mathbb{C}^3$ defined as follows \cite{Pawel}
\begin{equation}\label{RHO}
    \rho_a\, =\, \frac{1}{8a+1}\left(\begin{array}{ccc|ccc|ccc}
   a&\cdot&\cdot&\cdot&a&\cdot&\cdot&\cdot&a\\
   \cdot&a&\cdot&\cdot&\cdot&\cdot&\cdot&\cdot&\cdot\\
   \cdot&\cdot&a&\cdot&\cdot&\cdot&\cdot&\cdot&\cdot\\
   \hline
   \cdot&\cdot&\cdot&a&\cdot&\cdot&\cdot&\cdot&\cdot\\
   a&\cdot&\cdot&\cdot&a&\cdot&\cdot&\cdot&a\\
   \cdot&\cdot&\cdot&\cdot&\cdot&a&\cdot&\cdot&\cdot\\
   \hline
   \cdot&\cdot&\cdot&\cdot&\cdot&\cdot&b&\cdot&c\\
   \cdot&\cdot&\cdot&\cdot&\cdot&\cdot&\cdot&a&\cdot\\
   a&\cdot&\cdot&\cdot&a&\cdot&c&\cdot&b
   \end{array}\right) \ ,
\end{equation}
with
\begin{equation}\label{bc}
    b = \frac{1+a}{2}\ , \ \ \ \ c = \frac{\sqrt{1-a^2}}{2}\ ,
\end{equation}
where $a \in [0,1]$. The above matrix representation corresponds to
the standard computational basis $|ij\> = |i\> \ot |j\>$ in $\Cd\ot
\Cd$ and to make the picture more transparent we replaced all zeros
by dots. Since the partial transposition $\rho_a^\Gamma \geq 0$ the
state is PPT for all $a\in [0,1]$. It is easy to show that for $a=0$
and $a=1$ the state is separable and it was shown \cite{Pawel} that
for $a \in (0,1)$ the state is entangled. The entanglement of
(\ref{RHO}) was identified using so called range criterion
\cite{Pawel}. However, one may easily show that Horodecki state may
be detected also by the popular realignment criterion
\cite{R-1,R-2}. Actually, the family (\ref{RHO})  provides one of
the first examples of bound entanglement.

Recently,  Horodecki state was generalized for $\mathbb{C}^d \ot
\mathbb{C}^d$ \cite{PLA}.  Let us introduce $3 \times 3$ positive
matrix
\begin{equation}\label{X}
    X = b \, (|1\>\<1| + |d\>\<d|) + c\, (|1\>\<d| +
    |d\>\<1|) + a\, \sum_{k=2}^{d-1} |k\>\<k|\ ,
\end{equation}
and define $\rho_a$ as follows
\begin{equation}\label{RHO-d}
    \rho_a = \frac{1}{[d^2-1]a+1}\, \sum_{i,j=1}^d\, |i\>\<j|\ot \rho_{ij}\
    ,
\end{equation}
where
\begin{equation}\label{}
    \rho_{ii} = a \mathbb{I}_d\ , \ \ \ (i < d)\ , \ \ \rho_{dd} =
    X\ , \ \ \ \rho_{ij} = a |i\>\<j|\ , \ \ (i \neq j)\ .
\end{equation}
Clearly, for $d=3$ one recovers (\ref{RHO}). It was shown \cite{PLA}
that (\ref{RHO-d}) defines 1-parameter family of PPT states.
Moreover, for $0<a<1$ these state are entangled. Again it may be
easily shown using e.g. realignment criterion.

The aim of this Letter is to provide a huge generalization of
(\ref{RHO-d}). Actually, we provide $d$-parameter family of PPT
states and perform full separability/entanglement analysis. For
pedagogical reason we start with $d=3$ in the next section and
postpone the general construction for Section \ref{D}. Final
conclusions are collected in the last section.

\section{Generalized Horodecki-like states in $3 \ot 3$}

Consider the following 3-parameter family of states
\begin{equation}\label{RHO-G}
    \rho_3\, =\, {N_3}\left(\begin{array}{ccc|ccc|ccc}
   b_1&c_1&\cdot&\cdot&a&\cdot&\cdot&\cdot&a\\
   c_1&b_1&\cdot&\cdot&\cdot&\cdot&\cdot&\cdot&\cdot\\
   \cdot&\cdot&a&\cdot&\cdot&\cdot&\cdot&\cdot&\cdot\\
   \hline
   \cdot&\cdot&\cdot&a&\cdot&\cdot&\cdot&\cdot&\cdot\\
   a&\cdot&\cdot&\cdot&b_2&c_2&\cdot&\cdot&a\\
   \cdot&\cdot&\cdot&\cdot&c_2&b_2&\cdot&\cdot&\cdot\\
   \hline
   \cdot&\cdot&\cdot&\cdot&\cdot&\cdot&b&\cdot&c\\
   \cdot&\cdot&\cdot&\cdot&\cdot&\cdot&\cdot&a&\cdot\\
   a&\cdot&\cdot&\cdot&a&\cdot&c&\cdot&b
   \end{array}\right) \ ,
\end{equation}
where $b$ and $c$ are defined in (\ref{bc}), and
\begin{equation}\label{}
    b_k = a + \lambda_k(b-a)\ , \ \ \  c_k = \lambda_k c\ ,
\end{equation}
with $\lambda_1,\lambda_2\in [0,1]$ for $k=1,2$. Finally, the
normalization factor $N_3$ reads as follows
\begin{equation}\label{N3}
    N_3^{-1} = 8a+1  + (1-a)(\lambda_1 + \lambda_2)\ .
\end{equation}
It is clear that for $\lambda_1=\lambda_2=0$ it reduces to the
Horodecki state (\ref{RHO}). Let us observe that $\rho_3$ gives rise
to the direct sum decomposition
\begin{equation}\label{DS-1}
    \mathbb{C}^3 \ot \mathbb{C}^3 = \mathcal{H}_0 \oplus
    \mathcal{H}_{13} \oplus \mathcal{H}_{21} \oplus \mathcal{H}_{32}
    \ ,
\end{equation}
where
\begin{equation}\label{}
    \mathcal{H}_0 = \rm{span}_{\,\mathbb{C}} \{\, |11\>,\ |12\>,\ |22\>,\ |23\>,\ |33\>,\, |31\> \} \ ,
\end{equation}
and the remaining three 1-dimensional subspaces are defined as
follows
\begin{equation}\label{}
    \mathcal{H}_{13} = \rm{span}_{\,\mathbb{C}} \{\, |13\>\,\} \ ,\ \
\mathcal{H}_{21} = \rm{span}_{\,\mathbb{C}} \{\, |21\>\,\} \ ,\ \
\mathcal{H}_{32} = \rm{span}_{\,\mathbb{C}} \{\, |32\>\,\} \ .
\end{equation}
Hence the positivity of $\rho_3$  is governed by the positivity of
$6\times 6$ matrix $M_3$ written in the block form as follows
\begin{equation}\label{M3}
    M_3 = \left(\begin{array}{ccc} B_1 & A & A' \\ A^T & B_2 & A' \\
    A^{'T} & A^{'T} & B_3 \end{array}\right) \ ,
\end{equation}
with $2\times 2$ blocks given by
\begin{equation}\label{BAA}
    B_k = \left( \begin{array}{cc} b_k & c_k \\ c_k & b_k \end{array}\right) \
    , \ \ \ A= \left( \begin{array}{cc} a & 0 \\ 0 & 0 \end{array}\right) \
    , \ \ \ A'= \left( \begin{array}{cc} 0 & a \\ 0 & 0 \end{array}\right)
    \ \ ,
\end{equation}
where $b_3:= b$ and $c_3:=c$. Note, that $M_3 = {M}'_3 +
a|\phi_3\>\<\phi_3|$,  where $|\phi_3\> = |101001\> \in \mathbb{C}^2
\ot \mathbb{C}^3$ and ${M}'_3$ is block-diagonal with diagonal
blocks
\begin{equation}\label{B-tilde}
    \widetilde{B}_k = \lambda_k \left( \begin{array}{cc} b-a & c \\ c & b \end{array}\right)
    \ ,
\end{equation}
where $\lambda_3 := 1$. It is therefore clear that $M_3 \geq 0$ and
hence $\rho_3 \geq 0$ as well. Interestingly, its partial
transposition
\begin{equation}\label{RHO-G}
    \rho_3^\Gamma\, =\, {N_3}\left(\begin{array}{ccc|ccc|ccc}
   b_1&c_1&\cdot&\cdot&\cdot&\cdot&\cdot&\cdot&\cdot\\
   c_1&b_1&\cdot&a&\cdot&\cdot&\cdot&\cdot&\cdot\\
   \cdot&\cdot&a&\cdot&\cdot&\cdot&a&\cdot&\cdot\\
   \hline
   \cdot&a&\cdot&a&\cdot&\cdot&\cdot&\cdot&\cdot\\
   \cdot&\cdot&\cdot&\cdot&b_2&c_2&\cdot&\cdot&\cdot\\
   \cdot&\cdot&\cdot&\cdot&c_2&b_2&\cdot&a&\cdot\\
   \hline
   \cdot&\cdot&a&\cdot&\cdot&\cdot&b&\cdot&c\\
   \cdot&\cdot&\cdot&\cdot&\cdot&a&\cdot&a&\cdot\\
   \cdot&\cdot&\cdot&\cdot&\cdot&\cdot&c&\cdot&b
   \end{array}\right) \ ,
\end{equation}
gives rise to another direct sum decomposition
\begin{equation}\label{DS-2}
    \mathbb{C}^3 \ot \mathbb{C}^3 = \widetilde{\mathcal{H}}_{1} \oplus \widetilde{\mathcal{H}}_{2}
    \oplus \widetilde{\mathcal{H}}_{3}     \ ,
\end{equation}
where
\begin{eqnarray} \label{}
\widetilde{\mathcal{H}}_{1}   &=& \rm{span}_{\,\mathbb{C}} \{\, |11\>,\ |12\>,\ |21\>\ \} \ ,\nonumber \\
\widetilde{\mathcal{H}}_{2} &=& \rm{span}_{\,\mathbb{C}} \{\, |22\>,\ |23\>,\ |32\>\ \} \ ,\\
\widetilde{\mathcal{H}}_{3} &=& \rm{span}_{\,\mathbb{C}} \{\,
|33\>,\ |31\>,\ |13\>\ \} \ .\nonumber
\end{eqnarray}
Note that  $\rho_3^\Gamma \geq 0$ due to the positivity of three $3
\times 3$ matrices
\begin{equation}\label{M-tilde}
    \widetilde{M}_k =\left( \begin{array}{ccc} b_k & c_k & 0 \\ c_k & b_k & a \\ 0 & a & a \end{array}\right)
    \ , \ \ \ \ k=1,2,3\ ,
\end{equation}
where as before $b_3=b$ and $c_3=c$. Therefore, $\rho_3$ defines a
family of PPT states  parameterized by $a,\lambda_1,\lambda_2 \in
[0,1]$. Note, that for $a=0$ it reduces to the block-diagonal and
hence separable operator. For $a=1$ one has $b_k=a=1$ and $c_k=0$
and hence it reduces to the standard Horodecki state with $a=1$
which is known to be separable \cite{Pawel}. It turns out that
$\rho_3$ is entangled for $0 < a < 1$. This result is based on the
``PPT-symmetric extensions" by Doherty et al.
\cite{Doherty-1,Doherty-2}. Interestingly, the entanglement of
$\rho_3$ is only partially detected by the simple realignment
criterion \cite{R-1,R-2} (see the Fig. 1.) Note, that the standard
Horodecki state corresponding to $\lambda_1=\lambda_2=0$ is detected
by realignment. Other corners of the parameter square $[0,1]\times
[0,1]$ are detected as well.

\begin{figure}[t] \label{FIG}
\begin{center}
\epsfig{figure=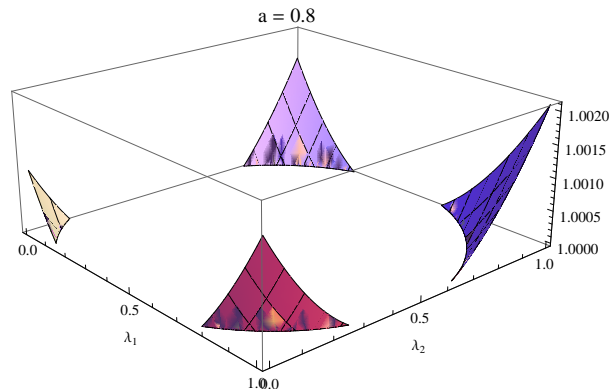,width=0.50\textwidth}
\end{center}
\caption{Realignment of $\rho_3$ for $a=0.8$. Note that only
``corners" of the parameter square $[0,1]\times [0,1]$ are detected.
}
\end{figure}

\section{Generalized Horodecki-like states in $d \ot d$} \label{D}

The above construction in $3 \ot 3$ may be easily generalized for $d
\ot d$ with arbitrary (but finite) $d$. Define $d \times d$ positive
matrix
\begin{equation}\label{X-lambda}
    X(\lambda) = b(\lambda) \, (|1\>\<1| + |d\>\<d|) + c(\lambda)\, (|1\>\<d| +
    |d\>\<1|) + a\, \sum_{k=2}^{d-1} |k\>\<k|\ ,
\end{equation}
with $b(\lambda)$ and $c(\lambda)$ being the following linear
functions of the parameter $\lambda\in [0,1]$
\begin{equation}\label{}
    b(\lambda) = a + \lambda (b-a)\ , \ \ \ c(\lambda) = \lambda\,
    c\ .
\end{equation}
Note, that $X(1)=X$, where $X$ was already defined in (\ref{X}). Let
\begin{equation}\label{}
    X_k = S^k X(\lambda_k) S^{k \dagger}\ ,
\end{equation}
where $S$ is the shift operator defined by
\begin{equation}\label{}
    S |k\> = |k+1\> \ , \ \ \ ({\rm mod}\ d)\ ,
\end{equation}
and $\lambda_k \in [0,1]$ for $k=1,\ldots,d$.  Finally, let us
introduce
\begin{equation}\label{}
    \rho_d = N_d \sum_{i,j=1}^d \, |i\>\<j| \ot \rho_{ij}\ ,
\end{equation}
where
\begin{equation}\label{}
    \rho_{ii} = X_i\ , \ \ \ \ \rho_{ij} = a\, |i\>\<j|\ , \ \ (i
    \neq j)\ .
\end{equation}
Fixing $\lambda_d=1$ one finds for the normalization factor
\begin{equation}\label{Nd}
    N_d^{-1} = [(d^2-1)a   + 1] + (1-a)\sum_{k=1}^{d-1}\lambda_k \ .
\end{equation}
Clearly, for $d=3$ this construction reproduces the previous one.
Note, that for $\lambda_1 = \ldots = \lambda_{d-1} = 0$ it
reproduces generalized Horodecki state from \cite{PLA}.

In analogy to (\ref{DS-1}) $\rho_d$  gives rise to the direct sum
decomposition
\begin{equation}\label{DS-d1}
    \mathbb{C}^d \ot \mathbb{C}^d = \mathcal{H}_0 \oplus
    \bigoplus_{k,l} \mathcal{H}_{kl}\ ,
\end{equation}
where
\begin{equation}\label{}
    \mathcal{H}_0 = {\rm{span}}_{\,\mathbb{C}} \{\, |ii\>,\ |i,i+1\>\ \} \
    , \ \ \ (i=1,\ldots,d\ \ {\rm mod}\ d) \ ,
\end{equation}
is $2d$-dimensional, and $d(d-2)$ 1-dimensional subspaces
\begin{equation}\label{}
    \mathcal{H}_{kl} = {\rm{span}}_{\,\mathbb{C}} \{\, |kl\>\,\} \ ,
\end{equation}
where the indices $k,l$ satisfy
\begin{equation}\label{ij-I}
k\neq l\ , \ \ \ l \neq k+1\ .
\end{equation}
Therefore, the positivity of $\rho_d$ reduces to the positivity of
$2d \times 2d$ matrix
\begin{equation}\label{}
    M_d = \sum_{i,j=1}^d |i\>\<j| \ot M_{ij}\ ,
\end{equation}
with
\begin{equation}\label{}
    M_{ii} = B_i\ , \ \ \ M_{ij} = A \ ,\  (i<j<d) \ , \ \ \ M_{id} =
    A'\ , \ (i<d) \ ,
\end{equation}
where the $2 \times 2$ matrices $B_i$, $A$ and $A'$ are defined in
(\ref{BAA}) (clearly, $i$ runs from $1$ up to $d$ and $b_d:=b$ and
$c_d:=c$). Note, that for $d=3$ one reproduces formula (\ref{M3})
for $M_3$. Now, the positivity of $M_d$ follows from the following
observation
\begin{equation}\label{}
    M_d = M'_d + a |\phi_d\>\<\phi_d|\ ,
\end{equation}
where $M'_d$ is block-diagonal with diagonal blocks
$\widetilde{B}_i$ defined in (\ref{B-tilde}) (with $\lambda_d=1$)
and $|\phi_d\> \in \mathbb{C}^2 \ot \mathbb{C}^d$ is defined by
\begin{equation}\label{}
    |\phi_d\> = (|10\> \oplus \ldots \oplus |10\>) \oplus |01\> \ ,
\end{equation}
where we have used $\mathbb{C}^2 \ot \mathbb{C}^d = \mathbb{C}^2
\oplus \ldots \oplus \mathbb{C}^2$ ($d$ terms).

Interestingly, the partial transposition $\rho_d^\Gamma$ is given by
\begin{equation}\label{}
    \rho^\Gamma_d = N_d \sum_{i,j=1}^d \, |i\>\<j| \ot \widetilde{\rho}_{ij}\ ,
\end{equation}
where
\begin{equation}\label{}
    \widetilde{\rho}_{ii} = \rho_{ii} = X_i\ , \ \ \ \ \widetilde{\rho}_{ij} = \rho_{ij}^T = a\, |j\>\<i|\ , \ \ (i
    \neq j)\ ,
\end{equation}
gives rise to another direct sum decomposition
\begin{equation}\label{DS-d2}
    \mathbb{C}^d \ot \mathbb{C}^d = \bigoplus_{i} \widetilde{\mathcal{H}}_i \oplus
    \bigoplus_{k,l} \widetilde{\mathcal{H}}_{kl}\ ,
\end{equation}
where there are $d$  subspaces which are 3-dimensional
\begin{equation}\label{}
    \widetilde{\mathcal{H}}_i = {\rm{span}}_{\,\mathbb{C}} \{\, |ii\>,\ |i,i+1\>,\ |i+1,i\> \} \
    , \ \ \ (i=1,\ldots,d\ \ {\rm mod}\ d) \ ,
\end{equation}
and $d(d-3)/2$ subspaces $\widetilde{\mathcal{H}}_{kl}$ which are
2-dimensional
\begin{equation}\label{}
    \widetilde{\mathcal{H}}_{kl} = {\rm{span}}_{\,\mathbb{C}} \{\, |kl\>,\ |lk\>\ \}
    \ ,
\end{equation}
where the indices $k,l$ satisfy
\begin{equation}\label{ij-II}
    k<l\ , \ \ l \neq k+1\ , \ \ k \neq l+1 \ , \ \ \ \ ({\rm mod}\
    d)\ .
\end{equation}
Equivalently, this condition may be formulated as follows:  given $
k \in \{1,\ldots,d-2\} $,  one has the following bound for $l$
\begin{eqnarray}\label{ij-III}
l = \left\{ \begin{array}{ll} k+2,\ldots,d-1\ , \ & {{\rm for}}\ \ k=1  \\
 k+2,\ldots,d\ , \  & {{\rm for}}\ \ k=2,\ldots,d-2\  \end{array} \right.  \ .
\end{eqnarray}
Note, that condition (\ref{ij-II}) is more restrictive that
(\ref{ij-I}).  For $d=3$ one has only 3-dimensional subspaces (the
set of indices $k,l$ satisfying (\ref{ij-II}) is empty) and hence
(\ref{DS-d2}) reduces to (\ref{DS-2}). Now, positivity of
$\rho_d^\Gamma$ is governed by the collection of $d$ $3\times 3$
matrices and `$d(d-3)/2$' $2\times 2$ matrices. It is easy to see
that all $2\times 2$ matrices are equal to $a|11\>\<11|$ which is
evidently positive, whereas $3\times 3$ matrices are nothing but
$\widetilde{M}_k$ defined by (\ref{M-tilde}) (where $i$ runs from 1
up to $d$ and $b_d=b$, $c_d=c$). Therefore, $\rho_d$ defines a
family of PPT states parameterized by $d$ parameters:
$a,\lambda_1,\lambda_2,\ldots,\lambda_{d-1} \in [0,1]$. Note, that
for $a=0$ it reduces to the block-diagonal and hence separable
operator. For $a=1$ one has $b_k=a=1$ and $c_k=0$ and hence it
reduces to the generalized Horodecki state with $a=1$ which is known
to be separable \cite{PLA}.

Let us introduce $d$ product vectors
\begin{equation}\label{}
    |\psi_k\>  = |k\> \ot \left( \sqrt{\frac{1-a}{2}}\, |k\> +  \sqrt{\frac{1+a}{2}}\,
    |k+1\> \right) \ , \ \ \ k=1,\ldots,d\ .
\end{equation}
One finds the following decomposition
\begin{equation}\label{}
    \rho_d = N_d   \, ( X_{\rm ent} + X_{\rm sep}) \ ,
\end{equation}
where
\begin{equation}\label{}
   X_{\rm sep} = \sum_{k=1}^d \lambda_k\, |\psi_k\>\<\psi_k|  \ ,
\end{equation}
with $\lambda_d=1$, and
\begin{equation}\label{}
    X_{\rm ent} =  a( dP^+_d +  Q_d) \ ,
\end{equation}
where $P^+_d$ denotes maximally entangled state and
\begin{equation}\label{}
    Q_d = \mathbb{I}_d \ot \mathbb{I}_d - \sum_{k=1}^d ( P_k \ot P_k + \lambda_k P_k
    \ot P_{k+1} ) \ ,
\end{equation}
with $P_k := |k\>\<k|$.  It is clear that $X_{\rm sep} $ is
separable and $X_{\rm ent}$ is entangled being an NPT operator.
Hence, $\rho_d$ is a convex combination of entangled and separable
states. Note that for $a=0$ the entangled part drops out and $\rho_d
 = N_d X_{\rm sep}$ with $N_d^{-1} = \sum_{k=1}^d \lambda_k$.
Again, using semi-definite programming based on the ``PPT-symmetric
extensions" by Doherty et al. \cite{Doherty-1,Doherty-2} we show
that for $0 < a < 1$ the state $\rho_d$ is entangled.

\section{Conclusions}

We constructed a rich $d$-parameter family of PPT sates in
$\mathbb{C}^d \ot \mathbb{C}^d$ and performed full
separability/entanglement analysis. These states generalize
Horodecki state in $\mathbb{C}^3 \ot \mathbb{C}^3$ \cite{Pawel} and
$\mathbb{C}^d \ot \mathbb{C}^d$ constructed recently in \cite{PLA}.
Interestingly, generalized Horodecki-like are invariant under the
action of unitaries of the following form
\begin{equation}\label{U}
    U = \Pi_0 + \sum_{k,l} e^{i\alpha_{kl}}\, \Pi_{kl}\ ,
\end{equation}
where the indices $k,l$ satisfy (\ref{ij-I}), and the projectors
$\Pi_0$, $\Pi_{kl}$ are defined as follows
\begin{equation}\label{}
    \Pi_0 = \sum_{k=1}^d P_k \ot (P_k +  P_{k+1}) \ ,\ \ \ \
    \Pi_{kl} = P_k \ot P_l\ .
\end{equation}
Note, that (\ref{U}) defines $d(d-2)$-dimensional commutative
subgroup of $U(d^2)$. The characteristic feature of (\ref{U}) is
that $U$ is nonlocal, that is, it cannot be written as $U_1 \ot U_2$
with $U_1,U_2 \in U(d)$. Therefore, the symmetry group of the
generalized Horodecki-like states have different symmetry than
states defined by (\ref{RHO-d}). It was shown \cite{PLA} that
(\ref{RHO-d}) is invariant under $U_{\bf x} \ot \overline{U}_{\bf
x}$, where
\begin{equation}\label{}
    U_{\bf x} = \sum_{k=1}^d e^{i x_k}\, P_k\ ,
\end{equation}
with $x_1=x_d$. Hence, in our generalized multi-parameter family the
local symmetry $U_{\bf x} \ot \overline{U}_{\bf x}$ is changed to
the nonlocal symmetry defined by (\ref{U}). The crucial difference
between local and nonlocal symmetries is related to he properties of
PPT states. Note, that if $\rho$ is invariant under $U_1 \ot U_2$,
that is $U_1 \ot U_2 \rho = \rho  U_1 \ot U_2$, then $\rho^\Gamma$
is invariant under $U_1 \ot \overline{U}_2$. No such simple relation
exists for nonlocal symmetries. In general even if $U \rho = \rho U$
there is no universal way to find the symmetry of $\rho^\Gamma$. It
turns out that in the case of generalized Horodecki-like states one
has $\widetilde{U} \rho^\Gamma = \rho^\Gamma \widetilde{U}$, where
$\widetilde{U}$ are initaries defined by
\begin{equation}\label{U}
    \widetilde{U} = \sum_{m=1}^d e^{i\beta_m}\, \widetilde{\Pi}_m +
    \sum_{k,l} e^{i\gamma_{kl}}\, \widetilde{\Pi}_{kl}\ ,
\end{equation}
where the indices $k,l$ satisfy (\ref{ij-II}), and the projectors
$\widetilde{\Pi}_m$, $\widetilde{\Pi}_{kl}$ are defined as follows
\begin{equation}\label{}
    \widetilde{\Pi}_m = P_m \ot P_{m} + P_m \ot P_{m+1} + P_{m+1} \ot P_m   \ ,\ \ \ \
    \widetilde{\Pi}_{kl} = P_k \ot P_l + P_l \ot P_k\ .
\end{equation}
Interestingly, generalized Horodecki-like entangled states with
local symmetry are detected by realignment criterion. In general it
is no longer the case for the states with nonlocal symmetry. These
states are detected in the full parameters range  by semi-definite
programming methods.

It would be interesting to construct a family of (indecomposable)
entanglement witnesses detecting the entanglement of generalized
Horodecki-like states in $\mathbb{C}^d \ot \mathbb{C}^d$.

\section*{Acknowledgments}

We thank Jacek Jurkowski for his help in plotting the Fig. 1.

\end{document}